# Ultraclean suspended graphene by radiolysis of adsorbed water

Hao Wang[†], Milinda Randeniya [†], Austin Houston [‡], Gerd Duscher[‡*], and Gong Gu[†*]

[†] Min H. Kao Department of Electrical Engineering and Computer Science, The University of Tennessee, Knoxville, TN 37916, USA

[‡] Department of Materials Science and Engineering, The University of Tennessee, Knoxville, TN 37916, USA

ABSTRACT:

Access to intrinsic properties of a 2D material is challenging due to the absence of a bulk that would dominate over surface contamination, and this lack of bulk also precludes effective conventional cleaning methods that are almost always sacrificial. Suspended graphene and carbon contaminants represent the most salient challenge. This work has achieved ultraclean graphene, attested by electron energy loss (EEL) spectra unprecedentedly exhibiting fine-structure features expected from bonding and band structure. In the cleaning process in a transmission electron microscope, radicals generated by radiolysis of intentionally adsorbed water remove organic contaminants, which would otherwise be feedstock of the notorious electron irradiation induced carbon deposition. This method can be readily adapted to other experimental settings and other materials, to enable previously inhibited undertakings that rely on the intrinsic properties or ultimate thinness of 2D materials. Importantly, the method is surprisingly simple and robust, easily implementable with common lab equipment.

TEXT

Access to intrinsic properties is requisite for fundamental inquiries into 2D materials. In an early example, the "universal" measured conductance of graphene samples was demystified and attributed to electron and hole puddles, partly due to charged impurities on graphene and in supporting substrates.[1,2] Later theoretical[3] and experimental works[4,5] established that suspended graphene, free from substrates that hosted the majority of charged impurities, approached in some aspects the intrinsic Dirac-point transport behavior, the accessibility of which was quantified.[6] The investigations into suspended graphene beyond electrical transport, however, have revealed the ubiquity of another type of contaminants, hydrocarbons, and the formidable challenges they pose.[7-11]

Transmission electron microscopy (TEM) in conjunction with microanalysis, capable of probing structures and properties simultaneously, is a favorite means for investigating suspended 2D materials. Although atomic-resolution images showing apparently clean graphene in nanometer-scale fields of view (FOVs) have become common display items in the literature, indications of contaminants invisible to microscopy have accumulated.[10-12] Graphene growth from existing edges, monitored by atomic-resolution scanning transmission electron microscopy (STEM) video, was attributed to invisible hydrocarbon contaminants as "free" precursors.[12] Investigations into the notorious electron beam-induced carbon deposition from such unintentional precursors have revealed their strong adhesion to and fast diffusion on graphene;[10,11,13] rendering these contaminant molecules irremovable by carefully conceived procedures and invisible to the highest-resolution microscopies, respectively. While the formidability of the challenge was further revealed by many audacious endeavors using specialized or advanced equipment, e.g. *in situ* nanomechanical scraping[14] and *in situ* or pre-situ high-temperature annealing,[9-11] limited improvements have been made over a three-stage cleaning procedure established more than a decade ago.[15]

These contaminants cast doubt on the existence of truly freestanding graphene, in light of possible roles played by the irremovable contaminants in stabilizing graphene,[7] even after the recognition[16] that out-of-plane undulations stabilize extended 2D structures. The contaminants have until now prevented the acquisition of electron energy loss (EEL) spectra of graphene with data quality comparable to those of graphite[17] to reflect the chemical bonding and band structure similarities.[18] This void of data hinders conclusive theoretical inquiries. Similarly in the low-loss region, the sensitivity to contaminants requires the spectrum of genuinely pristine graphene for further theoretical investigation.[7] The seemingly irremovable contaminants have also hampered applications exploiting the ultimate thinness of graphene.[10-13] Driven by such applications and the fundamental need to access intrinsic properties, the aforementioned long standing efforts have been directed towards clean graphene. While contamination is a concern common to all 2D materials due to the lack of a "bulk" that would dominate over "surface" contaminants, the issue is the most salient with graphene in the inevitable presence of hydrocarbon contaminants; obtaining clean graphene paves the way towards clean 2D materials in general.

In this work, we have achieved ultraclean graphene, attested by unprecedentedly resolved EEL spectra consistent with the graphene band structure and the chemical bonding in an ordered graphene crystal structure free of contaminants or defects. Importantly, our method is surprisingly simple and robust, readily implementable with common lab equipment and general-purpose transmission electron microscopes. The *in situ* cleaning method, based on the radiolysis reaction kinetics of hydrocarbons and water, can be adapted to other materials (not limited to 2D) and other experimental techniques.

The core-loss spectrum exhibits exceptionally high signal-to-background ratio, indicating ultra-cleanness because contaminants contribute to the background. We directly compare our raw data in Figure 1a with a graphite carbon-K (C-K) edge spectrum,[18] and further in Figure S1a with a graphite C-K energy loss near-edge structure (ELNES),[17] both among state-of-the-art graphite core-loss data. While also challenging, it is much easier to obtain graphite ELNES because the bulk dominates over surfaces even at typical TEM sample thicknesses, compared with graphene, a single atomic layer. Most noticeable, the separation between peaks B and C is unprecedentedly resolved. Here, we label the peaks A, B, C, etc. to confine this Letter in the experimental realm without delving into their bond and band origins, although the association of peak A to $\pi^*$ states is well known and other features also addressed[17, 18] in the literature.

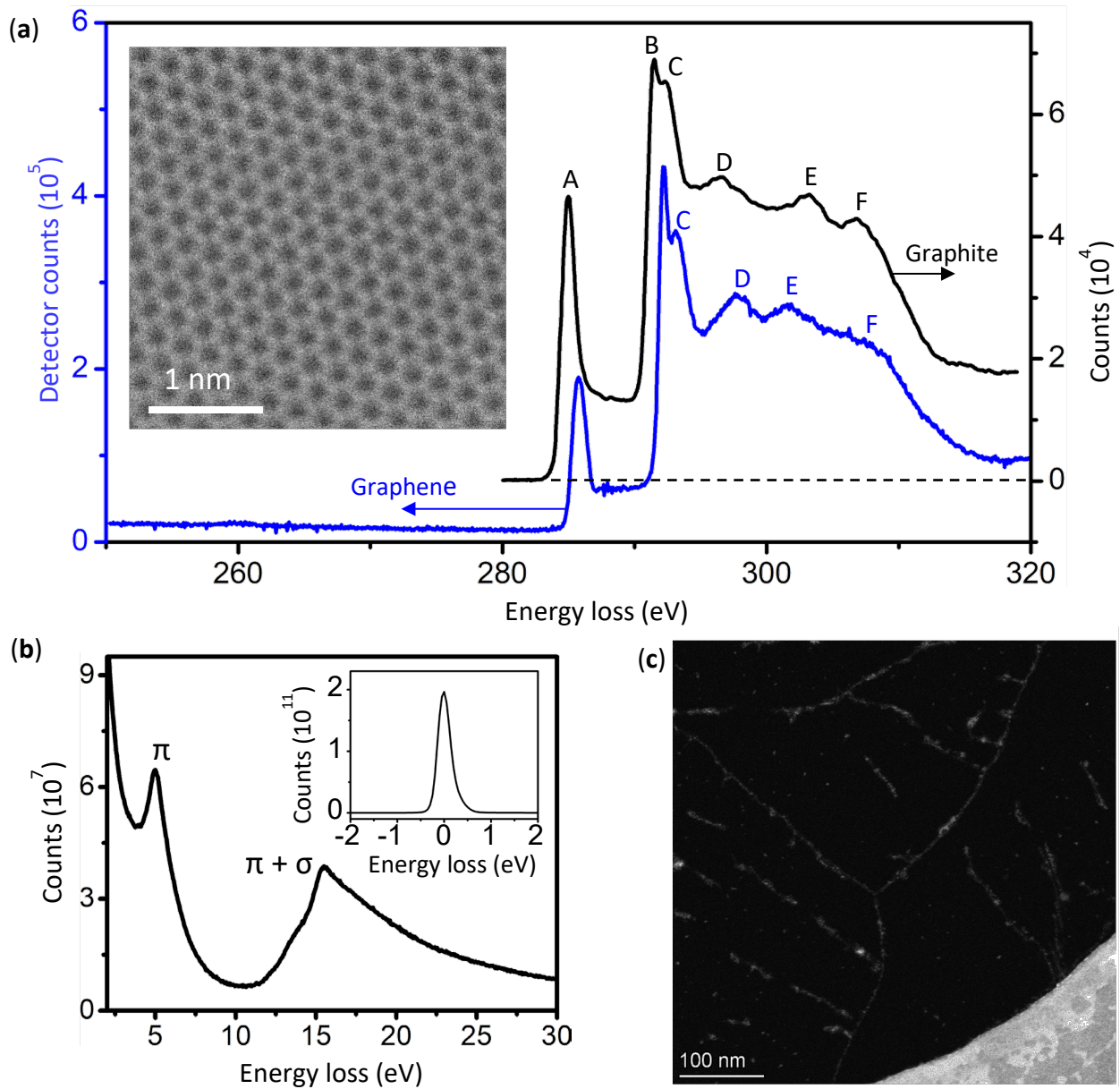

**Figure 1.** EEL spectra and STEM images of ultraclean graphene. Only raw data are shown in the main article to showcase data quality. (a) Core-loss spectrum of ultraclean graphene (in detector counts) in comparison with that of graphite,[18] each with zero count level indicated even though the latter is background-subtracted. For our graphene spectrum, ~16 detector counts represent one

detected electron, therefore each analyzer channel in the C-K spectral range detects a high electron count in the order of $10^4$ (acquisition details in Supporting Information). This raw spectrum is compared to energy resolution-sharpened ELNES of graphite,[17] analyzed to reveal peak positions and widths, and compared with state-of-the-art graphene core-loss spectrum raw data[19] in Figure S1 and its caption. Inset: Atomically resolved ADF STEM image acquired at 60 kV with high beam current (111 pA) and long exposure (11.5 minutes). (b) Low-loss EEL spectrum. Inset: Zero-loss peak. Details revealed by peak fitting are shown in Fig. S1. (c) Overview ADF STEM image acquired at 200 kV showing large clean areas with traces of solidified contaminants. Lower-magnification, larger-FOV overviews containing the area of this image are shown in Figure S2.

Cleanness is also key to achieving the unusually high signal-to-noise ratio of the spectrum, because the high signal is made possible by the extremely high accumulative dose (600 pA × 5 minutes); beam-induced carbon deposition from hydrocarbon contaminants limits the dose to many orders of magnitude lower on graphene samples prepared by previously developed methods. The luxury of high dose and long exposure was enjoyed in annular dark-field (ADF) STEM imaging (inset to Figure 1a) as well.

Also acquired with an unusually high dose, the raw low-loss EEL spectrum (Figure 1b) clearly reveals the π and π + σ plasmon peaks at 5.1 and 15.5 eV, respectively, modifying the earliest reported values (4.7 and 14.5 eV) extracted from noisy data[7, 20] and slightly updating or validating more recent results (4.90 and 15.4 eV).[21].

Large-area cleanness is far more important than the apparent cleanness in the nanometer-sized FOV of an atomic-resolution image. Figure 1c shows clean (dark) regions hundreds of nm in linear dimensions, with traces of solidified contaminants (gray) in between; these designations are explained in the Supporting Information and will become evident after we present the contamination and cleaning mechanisms.

The method for achieving the ultra-cleanness is simple and robust. Briefly, direct transfer[22] of chemical vapor deposition-synthesized graphene from Cu foil[23] onto a holey carbon covered TEM grid is followed by pre-cleaning in ambient air at 250 to 300 °C. The sample is then exposed to air of near-saturation humidity overnight. After sample insertion, an electron shower is applied, with parallel broad beam (typically 10 μm diameter) at 20 e$^-$/(Å$^2$ s) for 30 minutes, resulting in a level of cleanness that allows continuously operating a very intense STEM probe for an entire work day without carbon deposition. Both the shower time and the beam intensity can be varied in large ranges. The underlying principles center on the competition between hydrocarbon radiolysis, forming visible contaminants (traces in Figure 1c) by carbonization, and reaction of hydrocarbons with radicals generated from water radiolysis, removing the hydrocarbons by forming volatile products. The pre-cleaning, decreasing the hydrocarbon amount, and the water vapor exposure, ensuring sufficient water adsorption, skew the competition to the latter.

To clarify the role of adsorbed water, a sample was loaded into a microscope immediately after the 300 °C baking in air. Such samples are referred to hereafter as dry ones, and those loaded after extended moist air exposure as wet samples. The initial cleanness afforded by the clean transfer and pre-cleaning (see Supporting Information, as for all experimental details) is manifested by the large clean area spanning the 500 nm diameter blue circle with only a few specks (Figure 2a). This designation of regions of uniform contrast in such overview TEM images as "clean" areas has been long established.[8] We qualify this level of cleanness, at which a graphene area is free of contaminants visible to atomic-resolution (S)TEM, as "apparently clean". After the acquisition of Figure 2a, the circled area was subjected to 5 minutes of electron irradiation at 300 e⁻/(Å² s), a very high intensity relative to typical TEM illumination levels but many orders of magnitude below the $10^8$ to $10^9$ e⁻/(Å² s) level of focused beams for STEM in this work. The irradiation resulted in the formation of contaminants visible in the micrograph (Figure 2b), leaving apparently clean areas between the new specks rarely larger than 20 nm. Interestingly, an initially relatively "dirty" dry sample underwent much less contamination after the same irradiation (Figure 2c→d). The observations indicate that the visible contaminant specks (or ridges) form (or grow) by nucleation of diffusing precursors. Random nucleation occurs all over large clean areas (a→b), whereas small apparently clean areas hardly changed as the contaminant ridges enclosing them grew thicker and slightly wider (c→d), exhausting the invisible contaminants as precursors for the growth of existing ridges. The observations are consistent with the relative ease to obtain atomically clean STEM images of nm-scale regions encircled by solid visible contaminants, in comparison with large-area clean regions, as experienced by us and others but seldom mentioned in the literature.[10,11]

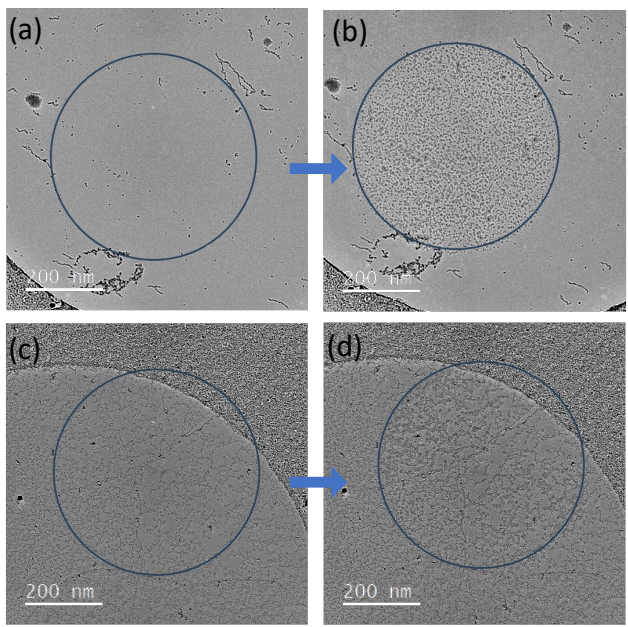

**Figure 2.** Beam-induced formation of visible contaminants. A clean dry sample suffers significantly more severe degradation in apparent cleanness (a→b) than does an initially dirty one (c→d), after the same 300 e⁻/(Å² s), 5 min irradiation. Both were loaded immediately after elevated-temperature pre-cleaning in ambient air.

Additional experiments (Figure S3) further clarified the dual role of electron irradiation. Briefly, the electron shower of our typical dose on dry samples is sufficient to largely prevent the formation of visible contaminants under moderate irradiance (typical of parallel illumination for TEM imaging), attributed to the unintentional presence of adsorbed water (or other species with similar effects). Nevertheless, the drastically more intense (~$10^8$ e⁻/(Å² s)) focused beam induces carbon deposition after the shower. Experiments further corroborating the effects of the water vapor exposure and electron irradiation are shown in Figures S4 and S5.

Intriguingly, the electron shower cleans graphene ≲ 5 μm beyond the showered area, manifested by the absence of carbon deposition upon scanning and dwelling a focused beam. The nonlocal cleaning is attributed to hydrocarbon diffusion into the showered area as a sink, strongly evidenced by the observation that the cleaning effect circumvented a ~0.8 μm wide void in suspended graphene (Figure S2).

By probing the roles played by water-deficient radiolysis and diffusion of hydrocarbons, we have demystified the notorious re-contamination of graphene samples. Here, the term re-contamination refers to the appearance of visible contamination in micrographs after apparent cleanness resulting from any treatment (e.g. pre-cleaning, in situ annealing) has been observed by imaging. Based on extensive experiments using dry samples (highlighted in Figure 3), we have concluded: 1) While graphene remains apparently clean under moderate electron irradiance (for TEM imaging), the radiolysis activated hydrocarbons are less mobile than the parent molecules and nucleate to form visible contaminants long after the beam is turned off.  2) Re-contamination by nucleation occurs only in irradiated areas and is insensitive to storage condition – little difference between ambient air and TEM vacuum, not surprising given the fast diffusion of hydrocarbons and the monolayer formation time[24,25] of seconds to minutes in typical TEM vacuum. 3) Moderately elevated temperatures promote surface diffusion but not desorption, therefore accelerate re-contamination.

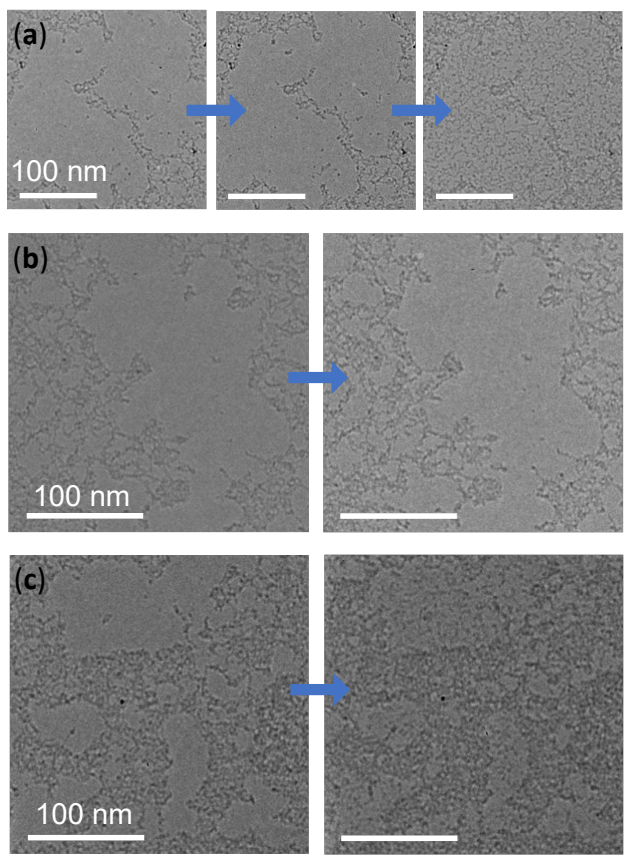

**Figure 3.** Re-contamination mechanisms, attributable to hydrocarbon radiolysis with insufficient water adsorption. All panels show TEM images of dry samples acquired with an illumination of 38 e$^-$/(Å$^2$ s), same as used for the electron shower for dry samples. The experiment shown in each row of images started in a region never previously irradiated, and the initial image was captured after 10 to 20 min of illumination. (a) Sequential TEM images of an area initially (left) exhibiting large apparently clean windows, remaining so (middle) after short *in situ* storage (10 min) with illumination blocked, but severely contaminated (right) after overnight *in situ* storage without illumination. The sample was kept at room temperature (RT). (b) An area was apparently clean (left) even after overnight storage in ambient air and remained so (right) after 15 min without illumination at RT. Comparing (b) with (a) reveals that irradiation prior to storage is the root cause of re-contamination regardless of the storage environment. (c) Another apparently clean area (left) after overnight ambient air storage was re-contaminated (right) after 15 min *in situ* storage without illumination at 90°C. Comparing (c) and (b) indicates that the raised temperature accelerates re-contamination after initial irradiation. All scale bars are 100 nm. The observations shown here are a subset of experiments presented in Figure S6 in greater detail.

Fast diffusion and radiolysis would not make the challenge so formidable without the worst culprit – the strong adhesion. While we have shown accelerated recontamination at 90 °C, others reported that much higher temperatures do not desorb the small-molecule contaminants.[9-11,13] A detailed study[10] showed that large-area apparently clean graphene can only be imaged by STEM at 800 °C while beam-induced carbon deposition still occurred at 500 °C, although atomic-level apparent cleanness is routinely shown by STEM in nm-scale regions encircled by visible contaminants.

From the observations by ourselves and others emerges a 2D nearly-ideal gas model for physisorbed small-molecule hydrocarbons on graphene surfaces, capturing both the strong adsorption and the high mobility in 2D, the latter despite the former due to a flat potential energy landscape of hydrocarbon molecules on graphene. Therefore, the molecules move fast to the irradiated spot or area as feedstock for beam-induced carbon deposition, rendering large-area clean graphene appear much dirtier than a nm-sized apparently clean window completely enclosed by solidified contaminants. This model is strongly supported by extensive theoretical and experimental studies. Molecular dynamics computations[26,27] revealed 2D gas behaviors of molecules, similar on graphene and graphite, and significantly higher diffusivity for $CH_4$ than inorganic molecules. Planar aromatics are the most studied,[28-32] conveying a picture of gliding platelets on a smooth landscape.

Strong adhesion renders physical desorption ineffective for cleaning and necessitates chemical means, for which adsorbed water radiolysis is the most convenient implementation in an electron microscope. In retrospect, the carbon deposition vs. hydrocarbon removal competition has been known in analytical electron microscopy (AEM)[33,34] of organic or biological specimens. Small hydrocarbon molecules are ubiquitous fast-moving contaminants on all surfaces. The mobile organic contaminants are dominated by hydrocarbons and are referred to as such here, leaving other possibilities (e.g. O-containing species) to be touched upon towards the end. When activated (i.e. dehydrogenated) by the electron beam in the absence of other chemical species, the hydrocarbons polymerize, cross-link, and eventually graphitize, manifested as visible contaminants, carbon deposition, or even epitaxial growth of graphene.[12] In the presence of species such as $H_2O$, $O_2$, and $N_2$, on the other hand, the radicals generated from their radiolysis react with hydrocarbons, activated or not, to form volatile products. Unintentional adsorption of $O_2$ and $N_2$ and residual water explains the cleaning effect of an electron shower on a dry sample.

In AEM, radicals from water radiolysis inevitably "etch" the organic or biological specimens and thus are not beneficial. The goal there is to avoid both net etching and net growth, rather than skew the competition. Here, water radiolysis products, in a very wide process window, removes hydrocarbons and leaves graphene intact, as attested by EEL spectra, thanks to the large chemical stability difference between graphene and the contaminants. In fact, we have not observed any defects (except tears and folds due to transfer) in atomic-resolution imaging in this work, whereas the same amount of imaging revealed many defects in our past work. While the underlying reasons are to be identified in future work, the graphene intactness after the 60 kV electron shower echoes previous work showing that the graphene Raman D band (defect signature) that emerges after irradiation by electrons below the knock-on damage threshold energy is solely associated with carbonaceous contaminants.[35]

The reaction kinetics principles are applicable to other processes. A portion of the final graphitic carbon residue originates, during our pre-cleaning, from the "incomplete combustion" of thick contaminant deposits due to the difficulty for $O_2$ to diffuse in. In light of this understanding, the second stage in the three-stage pre-cleaning,[15] vacuum annealing at 950 °C after thermal oxidation, further graphitizes the polymerized contaminants so that they do not fragment to re-contaminate the clean regions. The fixed residues formed during pre-cleaning or by hydrocarbon radiolysis in the TEM are visible in images, however, degrading the apparent cleanness. The third stage, extended vacuum annealing at 130 °C, is clean storage before TEM. Widely adopted variants of this procedure inevitably deprive graphene samples of beneficial adsorbed water, and so does *in situ* heating in the TEM.[10,11,36]

The identification of the visible contaminants as graphitic carbon residues resulting from hydrocarbon polymerization and graphitization are supported by C-K EEL spectra (Fig. 4). Here the term graphitic carbon refers to elemental carbon solids that are predominantly $sp^2$-coordinated regardless of order. The thin graphitic carbon is easily distinguishable from grown adlayers or bilayers formed by a graphene flake fallen on suspended graphene during transfer (Fig. S7). The spectrum of the thin residue highly resembles that of graphene, with peaks B and C clearly separated, each slightly broadened. The spectrum of the thick residue is as expected of graphitic carbon with disorders, exhibiting clear separation of the A and B (associated with $p_z$-π and $sp^2$-σ bonding), albeit merged B and C peaks. The broadening (thin and thick residues) indicates disorders, even though these features are better resolved than or as well as published graphene core-loss data.

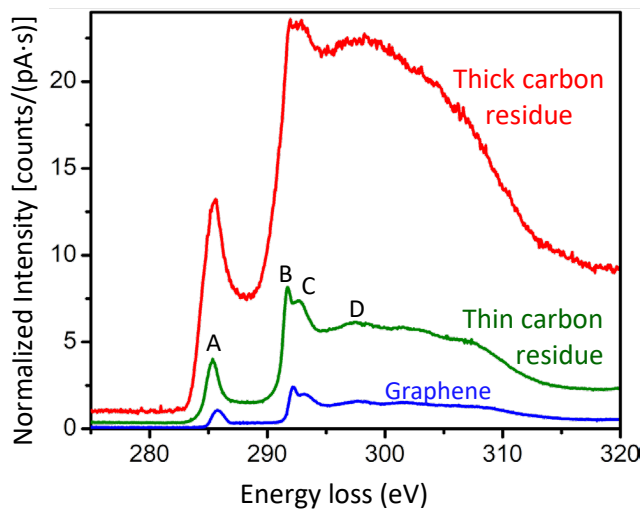

**Figure 4.** Normalized C-K edge EEL spectra of thick (red) and thin (green) graphitic residues compared with that of graphene (blue; same as in Figure 1a). The normalization against beam current and acquisition time is to facilitate comparison (see Supporting Information). The A peak height ratio is about 12 : 3.7 : 1 for thick, thin residues and bear graphene regions.

The ultraclean graphene is not perfectly clean, given the short monolayer formation time and fast hydrocarbon diffusion. By mass-contrast STEM imaging (Figure S8), we estimated the upper limit of the total mass density of invisible adsorbates on both sides to be ~4% of a single graphene layer.

The robust process works in a large parameter window (electron shower doses from $3.4 \times 10^4$ to $1.8 \times 10^5$ e$^-$/Å$^2$ explored) and at commonly used acceleration voltages from 60 to 300 kV.

Interestingly, a very low dose of 300 e$^-$/Å$^2$, 1/120 that of the standard shower, results in the same initial cleanness that nonetheless lasts only 40 min before carbon deposition occurs. Excess water condensed at cryogenic temperatures resulted in graphene damage (Figure S9), indicating that continuous water or ice films should be avoided.

As many experimental breakthroughs, this work points to more future endeavors than accomplished undertakings and poses more questions than it answers. First, there is plenty of room for the process itself to improve. The presumption that small-molecule organic contaminants are predominantly hydrocarbons is supported by the absence of O signatures in EEL spectra of our directly transferred[22] graphene (in contrast to the polymer residue originated from transfer).[14] Irremovable graphitic residues formed during the pre-cleaning can be decreased by limiting the thermal oxidation temperature, or even nearly eliminated by forgoing the step, with organic adsorbates containing groups such as –OH, –CO deliberately introduced to favor radiolysis reactions generating volatile products.

Second, the ultra-cleanness will remove the stubborn obstacle to the tantalizing observation of defects and heterostructures – "features" preferentially buried under contaminants. Direct, interpretation-free visualization of the atomic structure of the hexagonal boron nitride-graphene in-plane heterojunction by elemental-contrast STEM has not been successful,[37] and the challenge has stood unresolved despite the progress in synthesis.[38] This long sought revelation is now within reach. Similarly, the eradication of unintentional carbon deposition will allow the utilization of graphene as the ultimate thin support for applications such as focused electron beam induced deposition for direct write.[11,13]

Moreover, it is an enabler finally available for resolving some long-standing puzzles. The foremost fundamental question to answer may be how hydrophobic graphene is. In our process, enough adsorbed water remains on graphene in the TEM vacuum for a work day to fight off carbon deposition despite the intense STEM probe irradiating fast diffusing hydrocarbons from uncleaned areas, supporting the suggestion[39,40] that pristine freestanding graphene is not hydrophobic. The adsorbed water is included in the estimated total adsorbate amount (4% of the areal mass density) on the ultraclean graphene. The estimated total amount rules out their role in the stability of freestanding graphene, but theory and computations show that adsorbates are intricately related to the ripples responsible for the stability and to virtually all properties of graphene.[41,42] Ultraclean graphene will be the clean slate for experimenting with controlled adsorbates to sort out the relations. Furthermore, widely available electron beams in different energy ranges or other radiations can be used for the radiolysis based cleaning of supported graphene to investigate broader questions, such as the wettability of graphene on various substrates[43-47] and the mechanisms underlying remote epitaxy through graphene.[48]

## ASSOCIATED CONTENT

**Supporting Information**

Experimental details in Supporting Methods and additional data and analysis Supporting Figures (PDF)

## AUTHOR INFORMATION

**Corresponding Author**

\* Gong Gu – Min. H. Kao Department of Electrical Engineering and Computer Science, The University of Tennessee, Knoxville, TN 37916, USA; https://orcid.org/0000-0002-3888-1427; Email: ggu1@utk.edu

\* Gerd Duscher – Department of Materials Science and Engineering, The University of Tennessee, Knoxville, TN 37916, USA; https://orcid.org/0000-0002-2039-548X; Email: gduscher@utk.edu

**Author Contributions**

The manuscript was written through contributions of all authors. All authors have given approval to the final version of the manuscript.

**Funding Sources**

**Notes**

The authors declare no competing financial interest.

ACKNOWLEDGMENT

This work was partially supported by Volante Solutions Inc., by NIST through Award #60NANB21D179 (HW, MR, GG), and by the U.S. Department of Energy, Office of Science, Basic Energy Sciences, Materials Sciences and Engineering Division (AH, GD).

ABBREVIATIONS

2D, two-dimensional; EELS, electron energy loss spectroscopy; ELNES, energy loss near-edge structure; STEM, scanning transmission electron microscopy.

REFERENCES

1. Adam, S.; Hwang, E.; Galitski, V.; Das Sarma, S., A self-consistent theory for graphene transport. *Proceedings of the National Academy of Sciences* **2007,** *104* (47), 18392-18397. https://doi.org/10.1073/pnas.0704772104
2. Geim, A. K.; Novoselov, K. S., The rise of graphene. *Nature materials* **2007,** *6* (3), 183-191. https://doi.org/10.1038/nmat1849
3. Sarma, S. D.; Hwang, E., Density-dependent electrical conductivity in suspended graphene: Approaching the Dirac point in transport. *Physical Review B* **2013,** *87* (3), 035415. https://doi.org/10.1103/PhysRevB.87.035415
4. Bolotin, K. I.; Sikes, K. J.; Hone, J.; Stormer, H.; Kim, P., Temperature-dependent transport in suspended graphene. *Physical review letters* **2008,** *101* (9), 096802. https://doi.org/10.1103/PhysRevLett.101.096802
5. Du, X.; Skachko, I.; Duerr, F.; Luican, A.; Andrei, E. Y., Fractional quantum Hall effect and insulating phase of Dirac electrons in graphene. *Nature* **2009,** *462* (7270), 192-195. https://doi.org/10.1038/nature08522
6. Mayorov, A. S.; Elias, D. C.; Mukhin, I. S.; Morozov, S. V.; Ponomarenko, L. A.; Novoselov, K. S.; Geim, A.; Gorbachev, R. V., How close can one approach the Dirac point in graphene experimentally? *Nano letters* **2012,** *12* (9), 4629-4634. https://doi.org/10.1021/nl301922d
7. Gass, M. H.; Bangert, U.; Bleloch, A. L.; Wang, P.; Nair, R. R.; Geim, A., Free-standing graphene at atomic resolution. *Nature nanotechnology* **2008,** *3* (11), 676-681. https://doi.org/10.1038/nnano.2008.280
8. Algara-Siller, G.; Lehtinen, O.; Turchanin, A.; Kaiser, U., Dry-cleaning of graphene. *Applied physics letters* **2014,** *104* (15). https://doi.org/10.1063/1.4871997
9. Tripathi, M.; Mittelberger, A.; Mustonen, K.; Mangler, C.; Kotakoski, J.; Meyer, J. C.; Susi, T., Cleaning graphene: comparing heat treatments in air and in vacuum. *physica status solidi (RRL)–Rapid Research Letters* **2017,** *11* (8), 1700124. https://doi.org/10.1002/pssr.201700124
10. Dyck, O.; Kim, S.; Kalinin, S. V.; Jesse, S., Mitigating e-beam-induced hydrocarbon deposition on graphene for atomic-scale scanning transmission electron microscopy studies. *Journal of Vacuum Science & Technology B* **2018,** *36* (1). https://doi.org/10.1116/1.5003034
11. Dyck, O.; Lupini, A. R.; Rack, P. D.; Fowlkes, J.; Jesse, S., Controlling hydrocarbon transport and electron beam induced deposition on single layer graphene: Toward atomic scale synthesis in the scanning transmission electron microscope. *Nano Select* **2022,** *3* (3), 643-654. https://doi.org/10.1002/nano.202100188
12. Liu, Z.; Lin, Y.-C.; Lu, C.-C.; Yeh, C.-H.; Chiu, P.-W.; Iijima, S.; Suenaga, K., In situ observation of step-edge in-plane growth of graphene in a STEM. *Nature communications* **2014,** *5* (1), 4055. https://doi.org/10.1038/ncomms5055
13. Van Dorp, W. F.; Zhang, X.; Feringa, B. L.; Hansen, T. W.; Wagner, J. B.; De Hosson, J. T. M., Molecule-by-molecule writing using a focused electron beam. *ACS nano* **2012,** *6* (11), 10076-10081. https://doi.org/10.1021/nn303793w

14. Schweizer, P.; Dolle, C.; Dasler, D.; Abellán, G.; Hauke, F.; Hirsch, A.; Spiecker, E., Mechanical cleaning of graphene using in situ electron microscopy. *Nature Communications* **2020,** *11* (1), 1743. https://doi.org/10.1038/s41467-020-15255-3
15. Huang, P. Y.; Ruiz-Vargas, C. S.; Van Der Zande, A. M.; Whitney, W. S.; Levendorf, M. P.; Kevek, J. W.; Garg, S.; Alden, J. S.; Hustedt, C. J.; Zhu, Y., Grains and grain boundaries in single-layer graphene atomic patchwork quilts. *Nature* **2011,** *469* (7330), 389-392. https://doi.org/10.1038/nature09718
16. Meyer, J. C.; Geim, A. K.; Katsnelson, M. I.; Novoselov, K. S.; Booth, T. J.; Roth, S., The structure of suspended graphene sheets. *Nature* **2007,** *446* (7131), 60-63. https://doi.org/10.1038/nature05545
17. Batson, P., Carbon 1s near-edge-absorption fine structure in graphite. *Physical Review B* **1993,** *48* (4), 2608. https://doi.org/10.1103/PhysRevB.48.2608
18. Yuan, J.; Brown, L., Investigation of atomic structures of diamond-like amorphous carbon by electron energy loss spectroscopy. *Micron* **2000,** *31* (5), 515-525. https://doi.org/10.1016/S0968-4328(99)00132-8
19. Mohn, M. J.; Biskupek, J.; Lee, Z.; Rose, H.; Kaiser, U., Lattice contrast in the core-loss EFTEM signal of graphene. *Ultramicroscopy* **2020,** *219*, 113119. https://doi.org/10.1016/j.ultramic.2020.113119
20. Eberlein, T.; Bangert, U.; Nair, R.; Jones, R.; Gass, M.; Bleloch, A.; Novoselov, K.; Geim, A.; Briddon, P., Plasmon spectroscopy of free-standing graphene films. *Physical Review B* **2008,** *77* (23), 233406. https://doi.org/10.1103/PhysRevB.77.233406
21. Idrobo, J. C.; Zhou, W., A short story of imaging and spectroscopy of two-dimensional materials by scanning transmission electron microscopy. *Ultramicroscopy* **2017,** *180*, 156-162. https://doi.org/10.1016/j.ultramic.2017.02.002
22. Regan, W.; Alem, N.; Alemán, B.; Geng, B.; Girit, Ç.; Maserati, L.; Wang, F.; Crommie, M.; Zettl, A., A direct transfer of layer-area graphene. *Applied Physics Letters* **2010,** *96* (11). https://doi.org/10.1063/1.3337091
23. Li, X.; Cai, W.; An, J.; Kim, S.; Nah, J.; Yang, D.; Piner, R.; Velamakanni, A.; Jung, I.; Tutuc, E., Large-area synthesis of high-quality and uniform graphene films on copper foils. *science* **2009,** *324* (5932), 1312-1314. https://www.science.org/doi/10.1126/science.1171245
24. https://en.wikipedia.org/wiki/Monolayer.
25. Umrath, W., Fundamentals of vacuum technology. 1998. https://research.northeastern.edu/app/uploads/Vacuum_fundamentalspart1.pdf
26. Sun, C.; Bai, B., Gas diffusion on graphene surfaces. *Physical Chemistry Chemical Physics* **2017,** *19* (5), 3894-3902. https://doi.org/10.1039/C6CP06267A
27. Sun, C.; Bai, B., Diffusion of gas molecules on multilayer graphene surfaces: Dependence on the number of graphene layers. *Applied Thermal Engineering* **2017,** *116*, 724-730. https://doi.org/10.1016/j.applthermaleng.2017.02.002
28. Hedgeland, H.; Fouquet, P.; Jardine, A.; Alexandrowicz, G.; Allison, W.; Ellis, J., Measurement of single-molecule frictional dissipation in a prototypical nanoscale system. *Nature Physics* **2009,** *5* (8), 561-564. https://doi.org/10.1038/nphys1335
29. Ershova, O. V.; Lillestolen, T. C.; Bichoutskaia, E., Study of polycyclic aromatic hydrocarbons adsorbed on graphene using density functional theory with empirical dispersion correction. *Physical Chemistry Chemical Physics* **2010,** *12* (24), 6483-6491. https://doi.org/10.1039/C000370K

30. de Wijn, A. S., Internal degrees of freedom and transport of benzene on graphite. *Physical Review E* **2011,** *84* (1), 011610. https://doi.org/10.1103/PhysRevE.84.011610
31. Calvo-Almazán, I.; Bahn, E.; Koza, M.; Zbiri, M.; Maccarini, M.; Telling, M.; Miret-Artés, S.; Fouquet, P., Benzene diffusion on graphite described by a rough hard disk model. *Carbon* **2014,** *79*, 183-191. https://doi.org/10.1016/j.carbon.2014.07.058
32. Calvo-Almazán, I.; Sacchi, M.; Tamtögl, A.; Bahn, E.; Koza, M. M.; Miret-Artés, S.; Fouquet, P., Ballistic diffusion in polyaromatic hydrocarbons on graphite. *The Journal of Physical Chemistry Letters* **2016,** *7* (24), 5285-5290. https://doi.org/10.1021/acs.jpclett.6b02305
33. J.J. Hren, Barriers of AEM: contamination and etching, in J. Hren, Joseph I. Goldstein, David C. Joy (eds.), *Introduction to Analytical Electron Microscopy*, Springer 1979. https://doi.org/10.1007/978-1-4757-5581-7
34. Kohl, H.; Reimer, L., Transmission Electron Microscopy. *Springer Series in Optical Sciences* **2008,** *36*. Chapter 11, p. 468. https://catalogue.library.cern/literature/kj7a6-dp828
35. J.H. Choi, J. Lee, S. M. Moon, Y.-T. Kim, Hyesung Park, and Chang Young Lee, A Low-Energy Electron Beam Does Not Damage Single-Walled Carbon Nanotubes and Graphene. *J Phys Chem Lett* 7, 4739 (2016). https://doi.org/10.1021/acs.jpclett.6b02185
36. Islam, A. E.; Zakharov, D. N.; Carpena-Nuňez, J.; Hsiao, M.-S.; Drummy, L. F.; Stach, E. A.; Maruyama, B., Atomic level cleaning of poly-methyl-methacrylate residues from the graphene surface using radiolized water at high temperatures. *Applied Physics Letters* **2017,** *111* (10). https://doi.org/10.1063/1.5001479
37. Liu, L.; Park, J.; Siegel, D. A.; McCarty, K. F.; Clark, K. W.; Deng, W.; Basile, L.; Idrobo, J. C.; Li, A.-P.; Gu, G., Heteroepitaxial growth of two-dimensional hexagonal boron nitride templated by graphene edges. *Science* **2014,** *343* (6167), 163-167. https://doi.org/10.1126/science.1246137
38. Wang, H. S.; Chen, L.; Elibol, K.; He, L.; Wang, H.; Chen, C.; Jiang, C.; Li, C.; Wu, T.; Cong, C. X., et al. Towards chirality control of graphene nanoribbons embedded in hexagonal boron nitride. *Nature Materials* **2021,** *20* (2), 202-207. https://doi.org/10.1038/s41563-020-00806-2
39. Prydatko, A. V.; Belyaeva, L. A.; Jiang, L.; Lima, L. M.; Schneider, G. F., Contact angle measurement of free-standing square-millimeter single-layer graphene. *Nature communications* **2018,** *9* (1), 4185. https://doi.org/10.1038/s41467-018-06608-0
40. Zhang, J.; Jia, K.; Huang, Y.; Liu, X.; Xu, Q.; Wang, W.; Zhang, R.; Liu, B.; Zheng, L.; Chen, H., Intrinsic wettability in pristine graphene. *Advanced Materials* **2022,** *34* (6), 2103620. https://doi.org/10.1002/adma.202103620
41. Thompson-Flagg, R. C.; Moura, M. J.; Marder, M., Rippling of graphene. *Europhysics Letters* **2009,** *85* (4), 46002. https://doi.org/10.1209/0295-5075/85/46002
42. Sarma, S. D.; Adam, S.; Hwang, E.; Rossi, E., Electronic transport in two-dimensional graphene. *Reviews of modern physics* **2011,** *83* (2), 407. https://doi.org/10.1103/RevModPhys.83.407
43. Rafiee, J.; Mi, X.; Gullapalli, H.; Thomas, A. V.; Yavari, F.; Shi, Y.; Ajayan, P. M.; Koratkar, N. A., Wetting transparency of graphene. *Nature materials* **2012,** *11* (3), 217-222. https://doi.org/10.1038/nmat3228
44. Raj, R.; Maroo, S. C.; Wang, E. N., Wettability of graphene. *Nano letters* **2013,** *13* (4), 1509-1515. https://doi.org/10.1021/nl304647t
45. Li, Z.; Wang, Y.; Kozbial, A.; Shenoy, G.; Zhou, F.; McGinley, R.; Ireland, P.; Morganstein, B.; Kunkel, A.; Surwade, S. P., Effect of airborne contaminants on the wettability of supported graphene and graphite. *Nature materials* **2013,** *12* (10), 925-931. https://doi.org/10.1038/nmat3709

46. Hong, G.; Han, Y.; Schutzius, T. M.; Wang, Y.; Pan, Y.; Hu, M.; Jie, J.; Sharma, C. S.; Muller, U.; Poulikakos, D., On the mechanism of hydrophilicity of graphene. *Nano letters* **2016,** *16* (7), 4447-4453. https://doi.org/10.1021/acs.nanolett.6b01594

47. Ashraf, A.; Wu, Y.; Wang, M. C.; Yong, K.; Sun, T.; Jing, Y.; Haasch, R. T.; Aluru, N. R.; Nam, S., Doping-induced tunable wettability and adhesion of graphene. *Nano letters* **2016,** *16* (7), 4708-4712. https://doi.org/10.1021/acs.nanolett.6b02228

48. Kim, Y.; Cruz, S. S.; Lee, K.; Alawode, B. O.; Choi, C.; Song, Y.; Johnson, J. M.; Heidelberger, C.; Kong, W.; Choi, S., Remote epitaxy through graphene enables two-dimensional material-based layer transfer. *Nature* **2017,** *544* (7650), 340-343. https://doi.org/10.1038/nature22053

# Ultraclean suspended graphene by radiolysis of adsorbed water

Hao Wang[†], Milinda Randeniya [†], Austin Houston [‡], Gerd Duscher[‡*], and Gong Gu[†*]

[†] Min H. Kao Department of Electrical Engineering and Computer Science, The University of Tennessee, Knoxville, TN 37916, USA

[‡] Department of Materials Science and Engineering, The University of Tennessee, Knoxville, TN 37916, USA

## Supplementary Information

**Supplementary Methods**

    Graphene growth

    Graphene transfer and pre-cleaning

    Transmission electron microscopy and microanalysis

    The robust process

**Supplementary Figures**

    Figures S1 to S9

**References**

## Supplementary Methods

Experimental methods in this work include graphene growth, transfer and pre-cleaning, and transmission electron microscopy and microanalysis.

**Graphene growth**. We use graphene grown by low-pressure CVD on Cu foil (Alfa Aesar, item No.10950), resulting in a continuous graphene film that fully covers the Cu foil substrate and that is >95% monolayer with small patches of adlayers.[1] Briefly, the copper foil is cleaned by sonication baths in deionized water (DI water), acetone, isopropanol and a dilute hydrochloric acid solution (5%), each about 5 min. After a DI water rinse the foil is dried with a nitrogen stream. In a hot wall CVD system (OTF-1200x, MTI corporation), the copper foil is annealed at 1020 °C for 20 min, with a pressure ~ 100 mTorr sustained by a 25 standard cubic centimeters per minute (sccm) pure hydrogen flow, followed by the 20 min growth initiated by introducing 20 sccm methane to the gas flow and adjust the total pressure to ~280 mTorr. After growth, the sample is rapidly cooled down with the gas flows unchanged. The graphene grown in such processes is found to be polycrystalline with boundaries and the mean grain size is about 400 nm.[2]

**Graphene transfer and pre-cleaning**. We use two types carbon membrane covered Au grids – Quantifoil (400 mesh) and EMResolution (300 mesh), with periodically arranged regular holes and with irregular holes in the membranes, respectively. The grids are cleaned before use, by sequentially immersing in chloroform, acetone and isopropanol (2 hours each) at room temperature (RT). After natural drying, grids are annealed at 900 °C in pure Ar under atmosphere pressure. Graphene is transferred onto the cleaned grids by largely following the established direct transfer method.[3] The product of the transfer process is referred to as the sample, with graphene covering the carbon membrane and suspended over the through holes. The sample is pre-cleaned in a tube furnace at 250°C to 300°C for 30 min with the quartz tube open to ambient air.

**Transmission electron microscopy and microanalysis**. To demonstrate the universality of the method, we acquired micrographs using two transmission electron microscopes (TEMs) in a large range of conditions.

<u>FEI Spectra 300 MC</u> is a 300 kV S/TEM equipped with an energy filter and a monochromator, enabling a spatial resolution of ~0.5 Å in the STEM mode. Acceleration voltage options include 60, 200, and 300 kV. The energy resolution can reach 200 meV in electron energy-loss spectroscopy (EELS). The sample chamber vacuum is typical about $3 \times 10^{-8}$ Torr and no worse than $10^{-7}$ Torr. The EEL spectra in this work were obtained along with atomic-resolution ADF STEM images using Spectra 300 MC, referred to hereafter as Spectra.

<u>Zeiss Libra 200</u> is a 200 kV TEM/STEM with a spatial resolution of ~2.5 Å in the STEM mode. Referred to hereafter as Libra, it was used for many of the mechanistic studies, especially those on dry samples. The vacuum is $2 \times 10^{-7}$ to $2 \times 10^{-6}$ Torr.

<u>Extended moist air exposure</u>. To ensure sufficiently abundant adsorbed water for the radiolysis cleaning described below, the pre-cleaned sample is placed over DI water (without direct contact) in a covered beaker. The moist air exposure is typically overnight, after which the samples is loaded into a TEM. Experiments to reveal mechanisms using dry samples, i.e., those without intentional moist air exposure before loading, are clearly specified.

Electron shower in Spectra. After an area of interest is quickly located by TEM imaging at an illumination dose rate < 0.1 e−/(Å² s), an electron shower is performed. The standard shower in Spectra, using parallel illumination for 30 min at 20 e−/(Å² s) in a 10 μm diameter spot, guarantees > 7 hours continuous STEM operation using an intense focused beam without beam-induced carbon deposition. Varied doses used to explore the process window are clearly specified. The showered area is often subjected to scanning and dwelling focused beam tests in the STEM mode to verify the elimination of beam-induced carbon deposition. The beam current ~ 100 pA focused in a ~1 Å² spot corresponds to a dose rate of $6 \times 10^8$ e−/(Å² s). The dwelling time for the dwelling beam test is 1 min.

Electron shower in Libra. The standard shower in Libra is parallel illumination at 38 e−/(Å² s) for 20 min in a 10 μm diameter spot. The scanning and dwelling (1 min) focused beam tests use a dose rate of $8 \times 10^7$ e−/(Å² s), delivered by a 80 pA beam current focused in a ~6 Å² spot.

Imaging & spectroscopy in Spectra. TEM imaging in Spectra typically uses parallel illumination at dose rates of 0.1 to 1 e−/(Å² s) and exposure times from 2 to 5 seconds. Annular dark-field (ADF) STEM images are usually acquired with a high dose rate of $6 \times 10^8$ e−/(Å² s), delivered by a ~ 100 pA beam focused into ~1 Å² at a convergence semi angle of 30 mrad. The collection outer semi-angle is 200 mrad. Mass-contrast high-angle ADF (HAADF) images (at low magnifications, e.g. Figure 1c) uses an inner semi-angle of 73 mrad (at 200 kV) or 82 mrad (at 60 kV) with a camera length of 91 mm. Due to the mass contrast, clean graphene appears dark while solidified contaminants brighter in low-magnification HAADF STEM overview images, and the average electron count per pixel of an area in such an image can be used to estimate the mass density if the background (electron count / pixel of vacuum) is measured. For atom-resolution STEM, the inner semi-angle was reduced to 41 mrad (camera length 185 mm) to compensate for inadequate detector sensitivity. Therefore, the imaging mode is medium-angle ADF (MAADF). Although the acquired images (e.g. inset to Figure 1a) are not purely of element- or mass-contrast thus cannot be used to quantitatively estimate mass densities or identify elements, atoms appear bright while voids dark. All EEL spectra in this work were acquired in in the STEM mode by Spectra at 60 kV using a monochromator to achieve an energy resolution of 0.3 eV while the energy dispersion was 0.05 eV/channel. The monochromator was excited with a slit of 1 μm. The probe current was from 100 to 700 pA, and the exposure time was from 100 to 300 seconds. After aberration correction, the electron probe size was about 1 Å in diameter at low beam currents but degraded at high current to about 3 Å. The convergence and collection semi-angles were 40 and 50 mrad, respectively.

The core-loss spectrum in Figure 1a was acquired by scanning a 1 nm × 1 nm area over 5 minutes with a 600 pA focused beam, corresponding to an accumulated dose of $10^{12}$ e−. The exposure time of each frame was 5 seconds, and the spectrum was the sum 60 frames.

The atomic-resolution image displayed as the inset to Figure 1a is the sum of 10 stacked MAADF image frames, 1024 × 1024 pixels each, acquired at 111 pA beam current by the a 69.2 s frame time. The total exposure time is thus > 11.5 minutes.

The low-loss spectrum in Figure 1b is acquired at the same 1 nm × 1 nm area as the core-loss spectrum in Figure 1a, at the same beam current. The exposure time of each frame is 1 ms, orders of magnitude shorter than for the core-loss spectrum, since the intense zero-loss peak is included in this spectral region. The total integration time is 100 seconds. The zero-loss peak shows an energy resolution of 300 meV (FWHM, inset of Figure 1b).

The core-loss spectra of graphitic carbon residues in Figure 4 were acquired as follows: 700 pA beam current over a total integration time of 300 seconds and 100 pA for 100 seconds for the thin and thick graphitic residues (solidified contaminants), respectively. For fair comparison, Figure 4 shows each EEL spectrum in "normalized intensity", defined as detector counts normalized against the total dose used.

<u>Imaging in Libra</u>. TEM imaging in Libra was conducted with the same dose rate as the standard electron shower, 38 $e^-/(Å^2\ s)$, with exposure time from 1 to 2 seconds. HAADF STEM images were recorded during scanning and dwelling beam tests at a fixed focused beam intensity, $8 \times 10^7$ $e^-/(Å^2\ s)$.

**The robust process** works in a large parameter window and applies to commonly used electron energies. While the EEL spectra and atomic-resolution images in Figure 1 were acquired after the 20 $e^-/(Å^2\ s)$, 30 min standard shower (accumulative dose $3.6 \times 10^4$ $e^-/Å^2$), our experiments so far have shown that variations from 38 $e^-/(Å^2\ s)$ × 15 min = $3.4 \times 10^4$ $e^-/Å^2$ to 200 $e^-/(Å^2\ s)$ × 15 min = $1.8 \times 10^5$ $e^-/Å^2$ lead to the same outcome – intact graphene free of beam-induced carbon deposition for a work day of 1-minute dwelling beam tests and scanning imaging with a focused beam at least $8 \times 10^7$ $e^-/(Å^2\ s)$ in intensity. Interestingly, 1/120 of the standard shower dose, i.e., 1 $e^-/(Å^2\ s)$ × 5 min = 300 $e^-/Å^2$, results in the same initial outcome that nonetheless lasts only 40 min before carbon deposition occurs. Furthermore, the method works for commonly used acceleration voltages from 60 to 300 kV. Some experiments in varied conditions are shown in Figures S2 and S5. Figure S9 shows an example of excess water resulting in graphene damage, indicating that continuous water or ice films should be avoided.

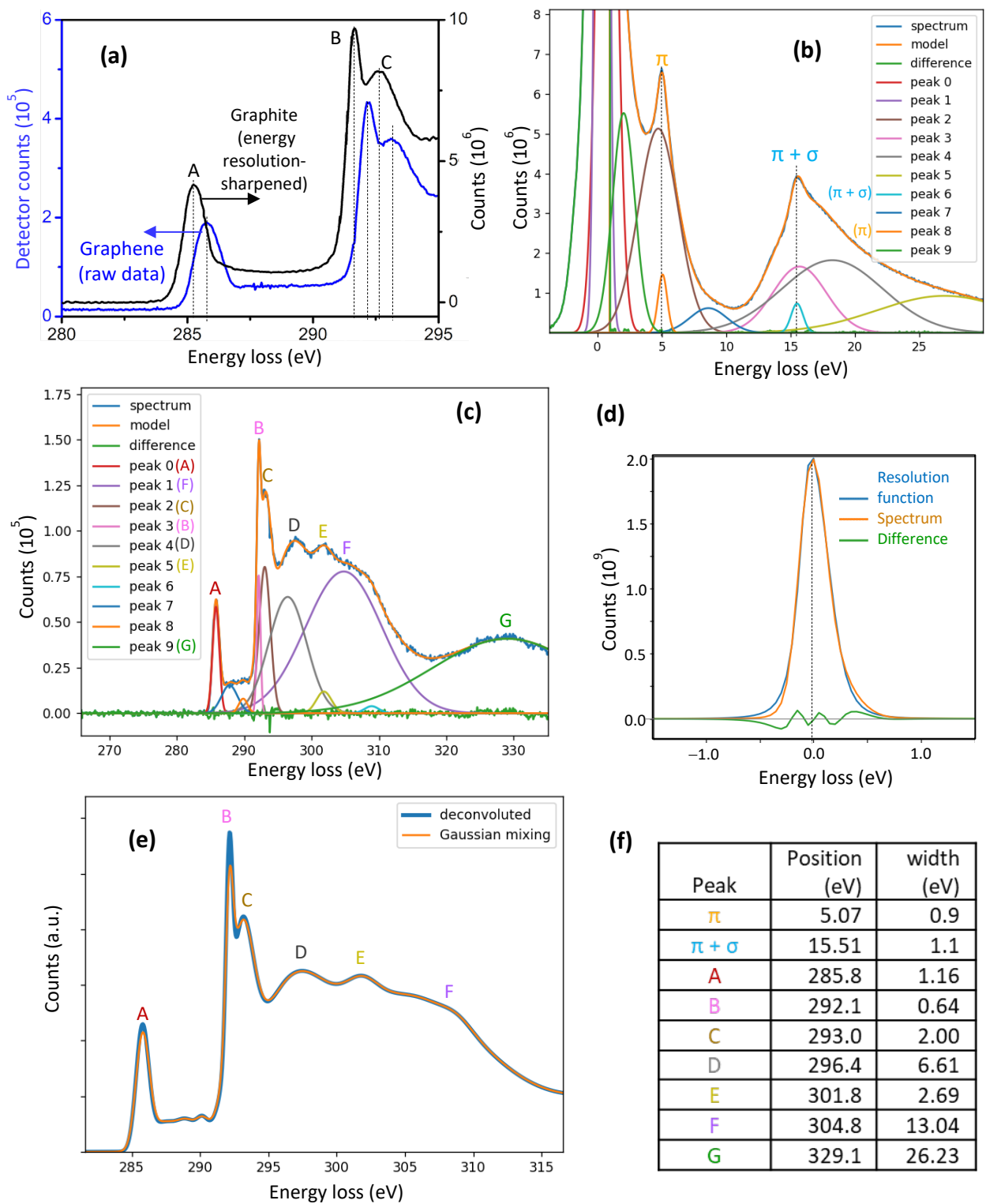

**Fig. S1.** EEL spectra of ultraclean suspended monolayer graphene.
(**a**) Raw C-K ELNES of ultraclean graphene (same spectrum as Figure 1a) in comparison with the energy resolution-sharpened, background-removed spectrum of a single-crystal domain of a 50-

80 nm in thick graphite sample.[4] Here, we use the graphite spectrum sharpened to a 0.17 eV energy resolution, while Figure 1 of the original paper[4] compares the results of sharpening to 0.17 and 0.22 eV resolutions, showing that the separation of peaks B and C is much more pronounced for the former. The effect of energy resolution function deconvolution on our graphene spectrum will be shown in (e). All peaks exhibit a rigid red shift from graphene to graphite, as expected, due to the shift between corresponding energy levels in the monolayer 2D and the bulk layered materials. The shift is also seen in Figure 1a. Also noteworthy is the unusually high signal-to-background ratio, quantified by a 16.4 : 1 ratio of the peak A maximum to the background at 280 eV. This is to be compared with the ratio of 2.3 : 1 for the graphene core-loss spectrum obtained by Mohn et al,[5] a rare instance of raw data without background subtraction and showing well-resolved peak A clearly separated from the takeoff of peak B. For our graphene spectrum, ~16 detector counts represent one detected electron, therefore each analyzer channel in the C-K spectral range detects a high electron count in the order of $10^4$, extremely high for a single atomic layer, leading to the unusually high signal-to-noise ratio. The high "signal" is made possible by the extremely high accumulative dose of $10^{12}$ e− (delivered by a 600 pA beam current over 5 minutes), given the low probability of inelastic scattering within the collection angle in the energy range of each channel.

(**b**) Peak fitting of a raw low-loss spectrum (without background subtraction) of the ultraclean graphene with Gaussian functions. The π and π + σ peaks are labeled. Peak 0 is the zero-loss peak. The curve labeled "model" is the sum of all Gaussians. The difference between the "model" and the raw spectra is displayed. A different spectrum than Figure 1b is used to demonstrate high reproducibility. This spectrum was acquired in the same sample area under the same conditions except a 10 s acquisition time (100 s for Fig. 1b).

(**c**) Peak fitting of a background-subtracted core-loss spectrum of the graphene. Peaks A through G are labeled, while ancillary peaks are also shown. The difference between the fitted (model) and the experimental spectra is displayed. The use of a different spectrum than Figure 1b (all conditions same except 100 s acquisition time here vs. 300 s) demonstrates high reproducibility.

(**d**) The zero-loss peak obtained by the fitting in (b), to be used as the energy resolution function. The raw spectrum and the difference are also shown.

(**e**) The energy resolution function-deconvoluted core-loss spectrum in comparison with the fitted experimental data (Gaussian mixing), showing only slight difference between the two. This is in contrast to the graphite data,[4] where considerable difference results from the sharpening to 0.17 and 0.22 eV energy resolutions. Here, for our graphene spectrum, we compare the fully energy resolution function-deconvoluted and the experimental spectra.

(**f**) List of energy positions and widths of characteristic peaks in (b) and (c).

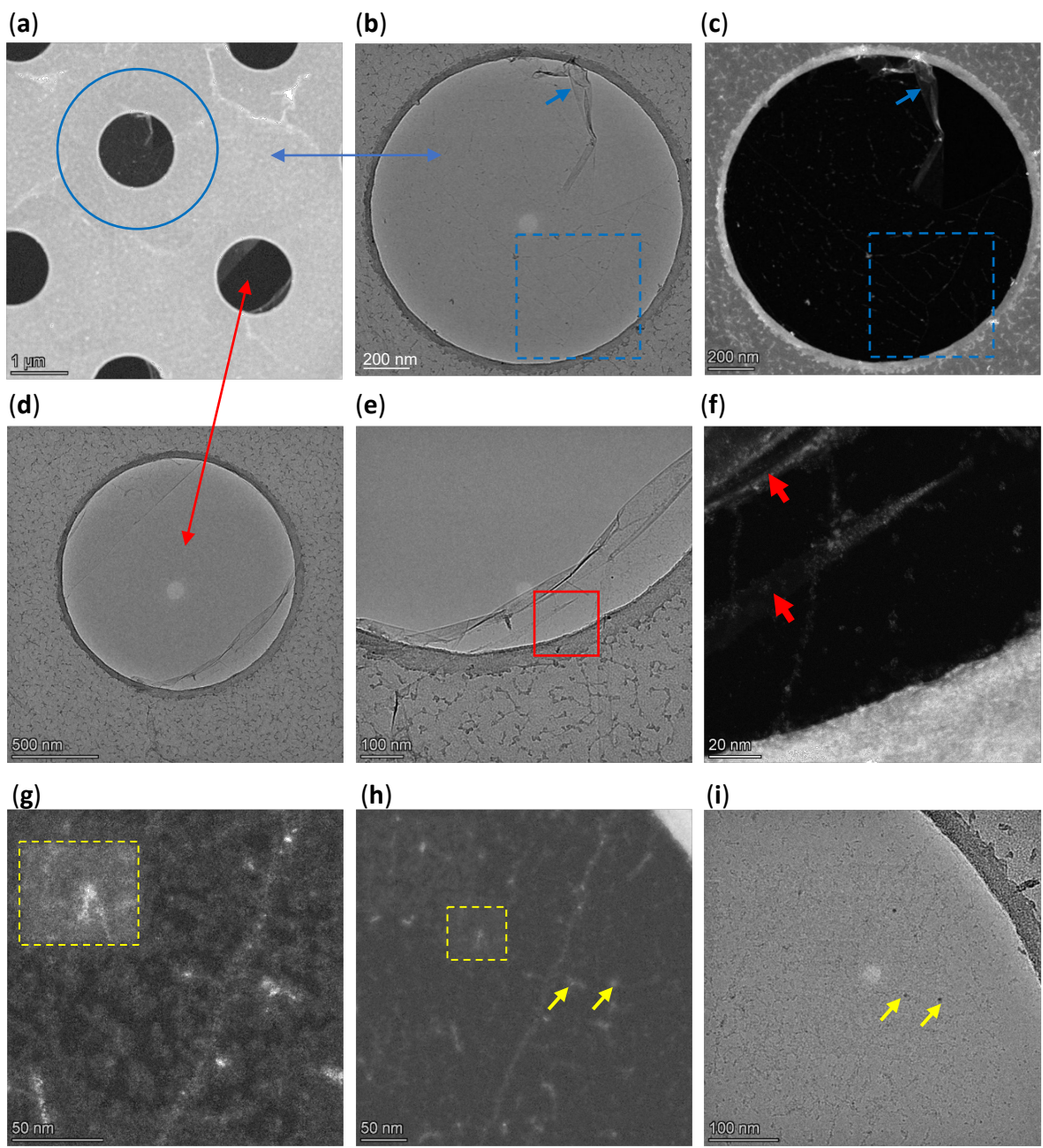

**Fig. S2.** Remote cleaning effect of electron shower.

(**a**) Overview of the showered (inside 3 μm diameter blue circle) and surrounding areas, acquired by HAADF STEM at 200 kV in Spectra (see Supplementary Methods above), as was Figure 1c in the main article. With the electron shower confined within a ~3 μm diameter circle, cleaning was effected ~2 μm away, on a sliver of graphene hanging from the edge of a carbon membrane hole and, in the direction towards the showered area, facing a ~0.8 μm wide vacuum gap resulting from the loss of suspended graphene during the transfer process.

(**b**, **c**) TEM and HAADF STEM images of graphene suspended over the carbon membrane hole inside the blue circle in (a), containing the area shown in Figure 1c. Indicated by a blue arrow is a sliver of broken graphene folded over, serving as a visible mark on the nearly featureless ultraclean graphene; thin traces of solidified contaminants can be discerned upon careful examination. The blue dashed boxes indicate the area imaged in Figure 1c. The bright spot in TEM image (b) is an artifact, as are those in all TEM images acquired by Spectra.

(**d**) TEM overview of the carbon membrane hole outside the showered area in (a), over which most of the graphene had been broken and lost. The sliver of suspended graphene across the void from the showered area was further examined to demonstrate remote cleaning.

(**e**) TEM images of the suspended sliver, showing folded graphene near the broken edge.

(**f**) STEM image of the red boxed area in (e). The absence of carbon deposition after raster scanning or dwelling of the intense focused beam manifests remote cleaning. The bright (signifying high mass thickness in HAADF STEM) traces indicated by red arrows are the folded-over graphene (rather than contaminants).

(**g**, **h**) In contrast to the remotely cleaned sliver in (d-f), STEM images of an area 130 μm away from the showered area, showing severe carbon depositions. The bright rectangle (in dashed box) is a previously raster scanned area.

(**i**) TEM image taken after the STEM images, showing the appearance of an area contaminated by beam-induced carbon deposition, with only small clean windows no larger than ~20 nm. The yellow arrows, as visual guides, indicate the same contamination spots in (h) and (i).

Images in all panels here as well as Figure 1c were taken at 200 kV in Spectra within a continuous session after the electron shower. These experiments also exemplify the wide applicability of the radiolysis-based *in situ* cleaning method to acceleration voltages higher than the graphene knock-on damage threshold (see also Figure S5) and to different shower schedules. The electron shower in the blue circle was at 200 e$^-$/(Å$^2$ s) for 15 min delivering an accumulative does of $1.8 \times 10^5$ e$^-$/Å$^2$, that is, 10 times the intensity, half the duration, and thus 5 times the total dose of the standard electron shower in Spectra.

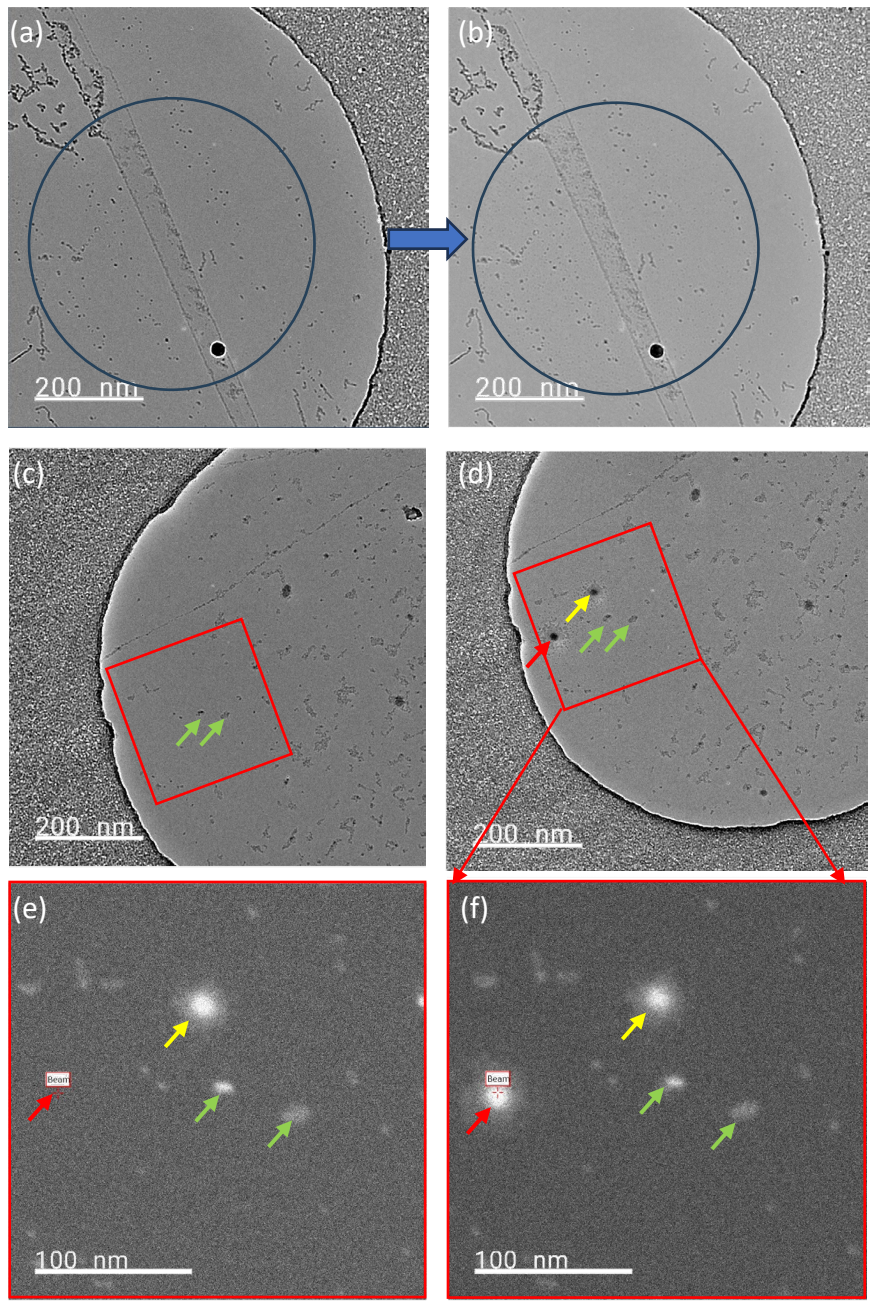

**Fig. S3.** Limited cleaning effect of electron shower on dry samples.

(**a,b**) TEM images of a dry sample before (a) and after (b) a 38 e−/(Å² s), 30 min shower. Following the shower, electron irradiation at 300 e−/(Å² s) for 10 min (the same dose rate as in the experiments in Figure 2 but double the duration) did not result in appreciable contamination, in contrast to Figure 2a→b. This experiment indicates that the shower, without preceding water vapor exposure, is sufficient to largely prevent visible contaminants under this irradiance.

(**c – f**) The cleanness achieved with the electron shower on a dry sample is insufficient to prevent carbon deposition induced by a very intense focused electron beam. Panels (c) and (d) are TEM images, respectively, before and after focused-beam irradiation tests. HAADF STEM image (e) of the boxed area in (c) and (d) was taken after a 1-minute dwelling-beam test, which resulted in the bright spot (carbon deposition) indicated by the yellow arrow. Two green arrows indicate pre-existing contamination spots. The red arrow indicates the location of the next dwelling beam test, after which image (f) was acquired, showing the resulting bright spot. The 1 min dwelling beam is a test for the cleanness, described in Supplementary Methods.

Images in all panels were taken at 200 kV in Libra.

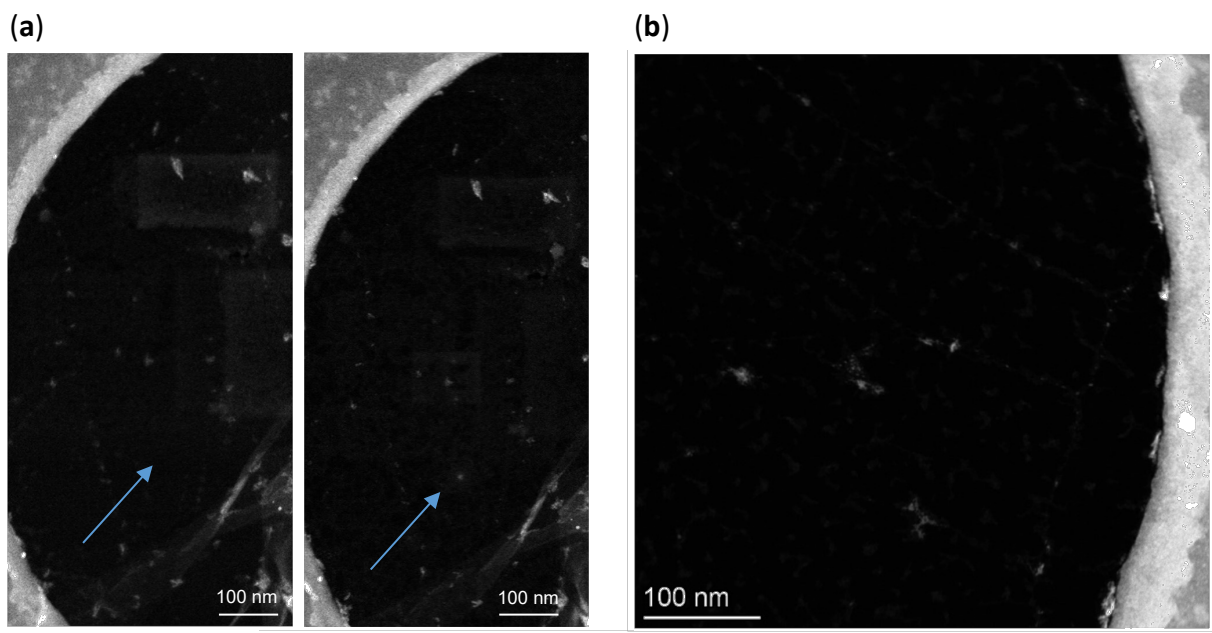

**Fig. S4.** The indispensable role of the electron shower. (**a**) Two STEM images sequentially taken in an un-showered area of a wet sample (intentionally exposed to moist air overnight), the right one after a 1 min dwelling beam test, showing a carbon contamination spot at the dwelling location marked by the arrows. The arrow in the left image indicates the corresponding location. White rectangles resulted from previous raster scans. (**b**) STEM image of a showered area after repeated raster scans and dwelling beam tests. No beam-induced carbon deposition was seen. This level of cleanness lasts a whole work day after the standard shower on a wet sample. All images taken at 60 kV in Spectra.

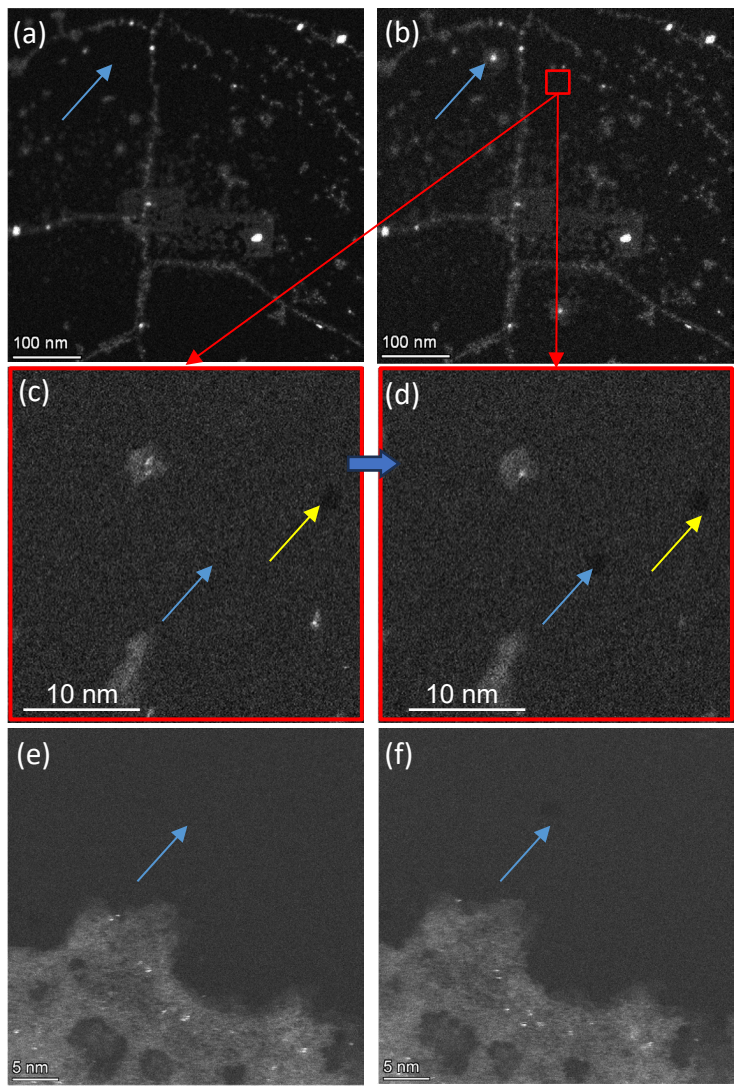

**Fig. S5.** Delayed electron shower is as effective as immediately following moist air exposure. Results also demonstrate that the cleaning method applies to commonly used acceleration voltages. (**a,b**) HAADF image acquired sequentially of a wet sample before shower. Gray rectangles are footprints of previous raster scans. Blue arrows indicate location of dwelling beam test, where bright spot in (b) manifests carbon deposition. (**c,d**) The boxed area in (b) after the standard electron shower is free of new beam-induced carbon deposition. In (c), the yellow arrow indicates a hole formed by a 1 min beam dwelling, while the blue arrow indicates the location for the next dwelling beam test, resulting a hole in image (d) taken after the test, due to the 200 kV voltage used, but the beam did not induce any carbon deposition. At this voltage, significantly higher than the knock-on damage threshold, the resulting point defects initiates etching of graphene by the water radiolysis generated radicals. Nevertheless, the graphene is intact upon raster scan imaging. (**e,f**) Similar effects of the standard shower at 300 kV acceleration voltage. After the electron shower, an intense focused beam does not induce any carbon deposition; while holes were drilled by the dwelling beam (location and a hole indicated by blue arrows), raster scanning leaves graphene intact. Images in all panels were taken in Spectra.

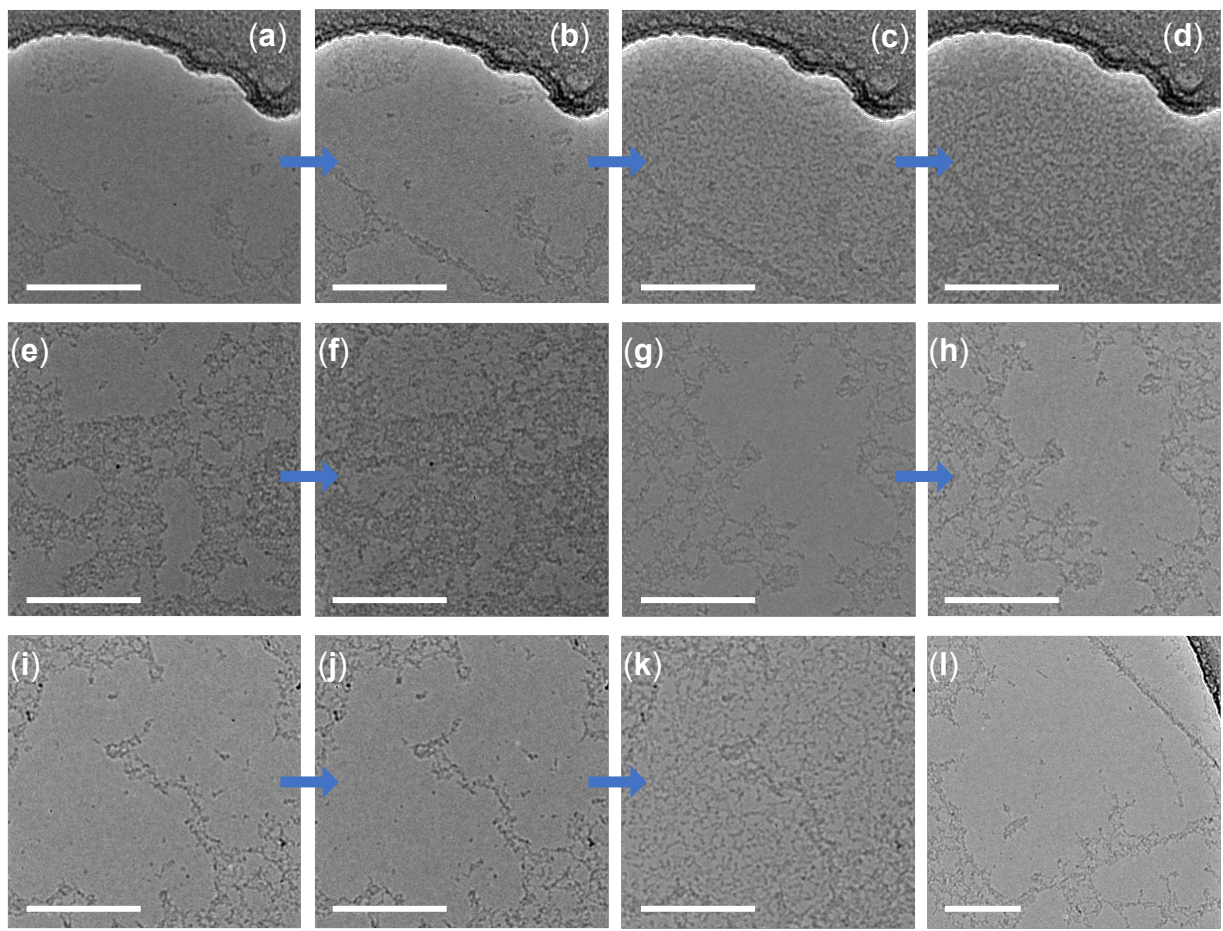

**Fig. S6.** Re-contamination and the role of sample temperature. All panels show TEM images of dry samples acquired at 200 kV in Libra under 38 e⁻/(Å² s) parallel illumination, same as used for the electron shower. All scale bars are 100 nm. Figure 3 in the main article show a subset of these images to highlight the major takeaways, while the experiments are shown here in detail and in chronological order.

(**a–d**) Irradiated area re-contaminated after air exposure. The initial image (a) was recorded after a pre-cleaned dry sample had been illuminated for 13 min. Similarly, all initial images in experiment sessions below were captured after 10 to 20 min of illumination. To study the effect of short-term *in situ* storage, image (b) was taken 15 min after the illumination was blocked, showing no sign of contamination. The sample was loaded back into Libra without pre-cleaning after overnight storage in ambient air, and the same area was imaged (c) immediately after being located, exhibiting severe re-contamination. Annealing the sample at 90°C *in situ* for 4 h without illumination worsens the degree of contamination (d).

(**e–h**) Un-irradiated area remains clean after air exposure. After the above 4 h annealing, initial image (e) of an area never previously illuminated exhibits a large clean window in the upper half. Following the initial illumination and a 15 min *in situ* dwelling with the sample temperature maintained at 90°C, the large clean window was severely contaminated. This is in contrast to the above 15 min *in situ* storage (a→b) at room temperature (RT). To clarify the role of temperature, another un-irradiated area was imaged (g) after cooling down to RT, and the large clean window was not re-contaminated (h) after a15 min *in situ* storage without illumination. These observations reveal that the root cause of re-contamination is previous irradiation and that a raised sample temperature accelerates re-contamination after the initial irradiation. Panels (e,f) show the same images as in Figure 3c of the main article, while images (g,h) are shown in Figure 3b.

(**i–l**) Re-contamination is insensitive to storage environment. An irradiated clean area of another dry sample (i) remained clean (j) after short-term (10 minutes) *in-situ* storage without illumination, but was re-contaminated (k) after overnight *in-situ* storage. Panels (i-k) show the same images as in Figure 3a. To confirm that the irradiation is the root cause of re-contamination (for this sample, as already shown for the above sample), image (l) of a previously un-irradiated area, taken after the same initial illumination and 15 min *in-situ* un-illuminated storage, exhibits no sign of re-contamination.

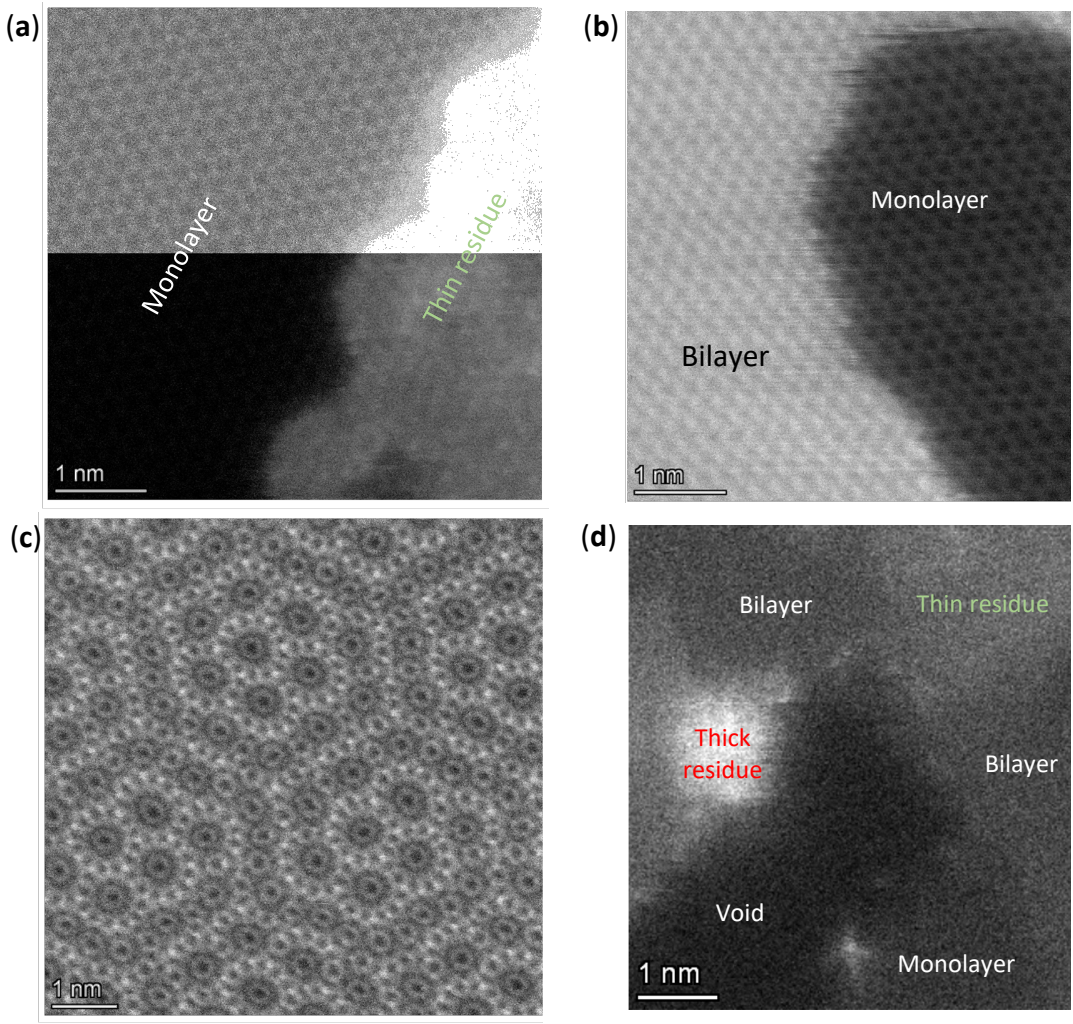

**Fig. S7.** Differentiate between graphitic carbon residues and graphene bilayers formed during growth or transfer. Thin graphitic carbon may be as thin as an equivalent graphene monolayer, with a total mass thickness of two graphene layers. Such graphitic carbon layers can be differentiated from bilayer graphene by atomic-resolution MAADF imaging. (**a**) Thin graphitic carbon (brighter than background) on suspended graphene. The upper and lower halves of the image are displayed with different gray scales to show both the graphene and the thin carbon residue with atomic-scale details. (**b**) Grown adlayer on monolayer graphene, exhibiting Bernal stacking order. (**c**) Twisted bilayer graphene exhibiting a moiré pattern, which may have been grown during the CVD process or formed by a broken-away graphene flake fallen on the suspended graphene during transfer. (**d**) An area with graphitic carbon and bilayer and monolayer graphene, as well as a void that provides a convenient measurement of the HAADF STEM image background to facilitate thickness estimations. The thin graphitic carbon region on bilayer graphene is estimated to be ~2 equivalent graphene layers thick, that is, a total of 4 equivalent graphene layers including the bilayer graphene underneath. The thick residue in the brown dashed contour is ~ 11 equivalent graphene layers thick (or a total of 13 layers including the bilayer graphene). All images were taken at 60 kV in Spectra.

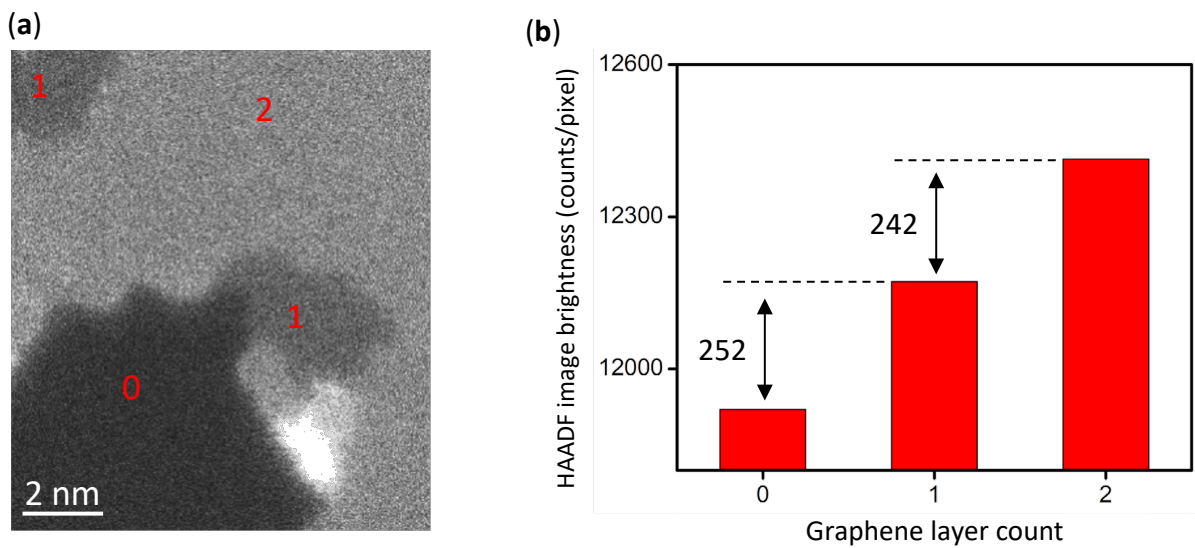

**Fig. S8.** Estimate the mass density of invisible adsorbates on ultraclean graphene. The mass-contrast HAADF STEM image (**a**, 60 kV, Spectra) shows regions of monolayer and grown bilayer regions, along with a void, convenient for background measurement. The chart (**b**) visualizes the average electron counts per pixel in these regions.

Let $M$ be the mass density of monolayer graphene and $m$ that of the invisible contaminants on one graphene surface, both in the unit of counts/pixel in the STEM image. For the monolayer regions, we have

$$2m + M = 252.$$

For the bilayer, we note that the amount of adsorbates on one surface is likely to be different for monolayer and bilayer suspended graphene but, for a rough estimate here, we assume that it is the same as on monolayer graphene. Therefore,

$$2m + 2M = 252 + 242.$$

Solving the above two equations, we get $2m/M \approx 4\%$.

Our experiments have suggested that the graphene surfaces retain adsorbed water for at least the entire work day. Therefore, water is included in the adsorbates, and the amount of organic contaminants is smaller than the estimate. Furthermore, the image was taken 2 h after the electrons shower, thus the initial amount of organic contaminates immediately after the shower may be significantly smaller (before contaminants diffuse from un-showered surrounding regions).

Therefore, we consider 4% of an equivalent graphene monolayer to be the upper limit of the total mass density of organic contaminants adsorbed on both surfaces of the ultraclean suspended graphene.

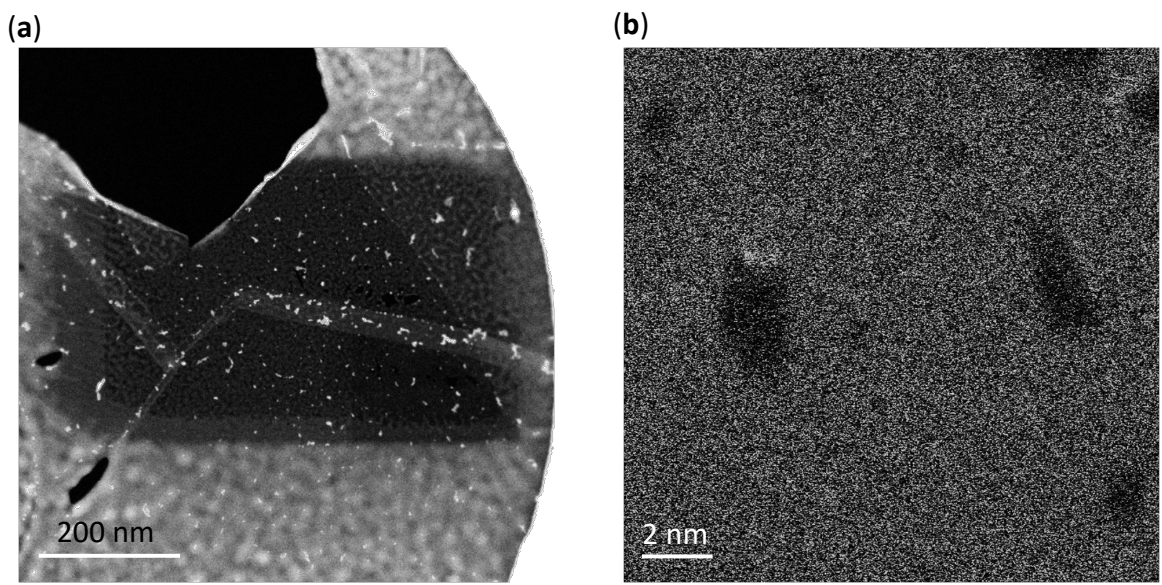

**Fig. S9.** Excess water at cryogenic temperature. The sample was loaded in a cryo-holder and the temperature was maintained at 100 K. (a) Low-magnification HAADF STEM image (60 kV, Spectra). The dark rectangle is the area of a previous raster scan, while the gray contrast (larger mass thickness) of the surrounding area is attributed to ice, condensed from residual water vapor in the vacuum. The dark contrast of the previously scanned area results from ice removal by radiolysis. The surrounding ice is ~ 5 equivalent graphene layers, estimated by electron counts (with the void facilitating background removal). (b) During raster scans for high-magnification STEM imaging, holes were generated and grew in size. Images (a) and (b) are acquired at a 40 pA beam current for total exposure times of 40 s and 1.5s, respectively, different from typical HAADF STEM imaging conditions in this work.

**Reference**

1. Li, X.; Cai, W.; An, J.; Kim, S.; Nah, J.; Yang, D.; Piner, R.; Velamakanni, A.; Jung, I.; Tutuc, E., Large-area synthesis of high-quality and uniform graphene films on copper foils. *science* **2009,** *324* (5932), 1312-1314.
2. Huang, P. Y.; Ruiz-Vargas, C. S.; Van Der Zande, A. M.; Whitney, W. S.; Levendorf, M. P.; Kevek, J. W.; Garg, S.; Alden, J. S.; Hustedt, C. J.; Zhu, Y., Grains and grain boundaries in single-layer graphene atomic patchwork quilts. *Nature* **2011,** *469* (7330), 389-392.
3. Regan, W.; Alem, N.; Alemán, B.; Geng, B.; Girit, Ç.; Maserati, L.; Wang, F.; Crommie, M.; Zettl, A., A direct transfer of layer-area graphene. *Applied Physics Letters* **2010,** *96* (11).
4. Batson, P., Carbon 1s near-edge-absorption fine structure in graphite. *Physical Review B* **1993,** *48* (4), 2608.
5. Mohn, M. J.; Biskupek, J.; Lee, Z.; Rose, H.; Kaiser, U., Lattice contrast in the core-loss EFTEM signal of graphene. *Ultramicroscopy* **2020,** *219*, 113119.